\documentclass[referee]{raa}

\usepackage{graphicx,times}
\usepackage{natbib}
\usepackage{lineno}
\usepackage{amssymb,amsmath}
\bibpunct{(}{)}{;}{a}{}{,}

\usepackage[pagebackref=true]{hyperref}

\begin{document}


\title{A New Pathway to Single Be Stars: Ejected Companions from Type Ia Supernovae}

\volnopage{Vol.0 (20xx) No.0, 000--000}
\setcounter{page}{1}

\author{Yuchen Bao
 \inst{1,2,3}
\and Zhenwei Li
 \inst{1,3}
 \thanks{lizw@ynao.ac.cn}
\and Hongwei Ge
 \inst{1,3}
\and Hailiang Chen
 \inst{1,3}
 \and Dengkai Jiang
 \inst{1,3}
\and Xuefei Chen
 \inst{1,2,3}
 \thanks{cxf@ynao.ac.cn}
\and Zhanwen Han
 \inst{1,2,3}
}

\institute{Yunnan Observatories, Chinese Academy of Sciences (CAS), Kunming 650216, People’s Republic of China; {\it baoyuchen@ynao.ac.cn}\\
\and
School of Astronomy and Space Science, University of Chinese Academy of Sciences (UCAS), Beijing 100049, People’s Republic of China\\
\and
International Centre of Supernovae (ICESUN), Yunnan Key Laboratory, Kunming 650216, People’s Republic of China\\
\vs\no
{\small Received 20xx month day; accepted 20xx month day}}

\abstract{Be stars are rapid rotators generally produced by binary interactions. The single Be stars in the observations pose challenges to the Be star formation theory. In this paper, we propose a new pathway for the formation of single Be stars, in which the Be star is taken as the ejected companion star from a Type Ia supernova (SN Ia) explosion. Recent numerical simulations suggest that explosive oxygen burning, initiated via the convective Urca process in certain helium (He) stars near the Chandrasekhar mass limit, can set off a SN Ia. Based on this proposition, we further demonstrate that about $0.4\%$ of He star + main-sequence (MS) star binaries may evolve into single Be stars, where the MS star is spun up due to the mass accretion from the He star, and then the He star explodes as a SN Ia. We employ detailed binary evolutionary simulations and find the parameter space that would produce single Be stars via the SN Ia channel. Around $22\%$ of Be stars from the SN Ia progenitor channel exhibit peculiar tangential velocities exceeding $24\ \rm km/s$, classifying them as runaway stars. This suggests that the SN Ia channel plays a meaningful role in forming single Be stars, particularly within the runaway star population.
}

\authorrunning{Yuchen Bao, Zhenwei Li, Hongwei Ge, Hailiang Chen, Dengkai Jiang, Xuefei Chen, Zhanwen Han}
\titlerunning{Single Be star as SN Ia remnant}

\maketitle

\section{Introduction} 
\label{sect:intro}

\citet{1866AN.....68...63S} first discovered the class of stars now known as Be stars. Since then, the studies of Be stars have advanced the understanding of the universe, star evolution, mass transfer, and binary evolution. Be stars are a class of rapidly rotating main-sequence (MS) stars that exhibit H$\alpha$ emission lines in their spectra (e.g., \citealp{1991A&A...241..419P,1997A&A...318..443Z,1998A&A...338..505N,2000ASPC..214....1B,2001A&A...368..912Y,2003PASP..115.1153P,2005ApJS..161..118M,2008A&A...478..467E,2010ApJ...722..605H,2013A&ARv..21...69R,2016A&A...595A.132Z,2024arXiv241106882R}). The circumstellar disk surrounding the central star is generally believed to be the origin of emission features (e.g., \citealp{1931ApJ....73...94S,1988PASP..100..770S,1991MNRAS.250..432L,1996A&A...308..170H,1997ApJ...477..926W,2001PASJ...53..119O,2003PASP..115.1153P,2006ApJ...639.1081C,2009A&A...504..915C,2009MNRAS.392..383J,2009ApJ...699.1973S,2011IAUS..272..325C,2013A&A...553A..25G,2013ApJS..204....2M,2013A&ARv..21...69R,2019MNRAS.488..387B}). The unique spectral features of Be stars make them important in astrophysics, as studying their spectral lines helps us understand physical processes like accretion and mass loss.

Previous works suggested that more than $75\%$ Be stars are formed via binary interaction \citep{2005ApJS..161..118M}. Therefore, Be stars are generally considered products of binary interaction, and studying their formation can effectively constrain the stability and efficiency of the transfer process (e.g., \citealp{1982IAUS...98..327R,1991A&A...241..419P,2013ApJ...764..166D,2014ApJ...782....7D,2013A&ARv..21...69R,2014ApJ...796...37S,2020ApJ...898..143S,2018MNRAS.477.5261B,2020A&A...638A..39L,2021A&A...653A.144H,2024A&A...683A..37Y,2025A&A...697A.239H,2026arXiv260206259X}). In a binary evolutionary scenario, when a star fills its Roche lobe, the donor transfers mass and angular momentum to the accretor through the inner Lagrangian point, resulting in stable mass transfer. As the accretor gains mass and angular momentum, it spins up. A classical Be star is formed once the accretor is spun up to near-critical rotation \citep{2004MNRAS.350..189T,2006ApJ...639.1081C,2010ApJ...722..605H,2013A&ARv..21...69R}. Within the binary evolution channel, Be star formation is associated with two primary types of mass transfer processes: the MS star channel and the helium (He) star channel. The MS star channel has been investigated in several studies \citep{1991A&A...241..419P,1995A&A...296..691P,1997A&A...322..116V}. A binary system consists of two MS stars, where the more massive MS star first fills its Roche lobe and transfers material to the less massive MS star. If the mass transfer is stable and avoids a common envelope (CE) stage, the companion accretes material, spins up, and ultimately forms a Be star. The stripped star will evolve to be a He-burning star and eventually become a compact object, such as a white dwarf (WD), neutron star (NS), or black hole (BH) \citep{2002MNRAS.331.1027D,2015MNRAS.451.2123T,2021A&A...645A...5S,2022A&A...661A..60A}. In contrast, the He star channel involves a binary system consisting of a MS star and a He star \citep{2009ApJ...707..870B}. In this case, the He star evolves and fills its Roche lobe, transferring material to the MS star. Through stable mass transfer, this process leads to the formation of a Be star. Alternatively, the merger of two MS stars in a contact binary may also form a Be star, which is known as the merger channel \citep{2013ApJ...764..166D,2013MNRAS.428.1218J,2014ApJ...796...37S}.

In addition to the binary evolutionary scenario, a Be star can also originate from the single-star channel, where spinning up through the gain of angular momentum from the natal molecular cloud or the transport of angular momentum from the contracting core to the outer envelope (e.g., \citealp{2000A&A...361..101M,2008A&A...478..467E,2008MNRAS.388.1879M,2020A&A...633A.165H,2022A&A...667A.111K}).

Observations reveal that many Be stars appear as isolated objects \citep{2001ApJ...555..364B,2020A&A...641A..42B,2022ApJS..260...35W,2023A&A...679A.109C}. The most natural formation channels for such single Be stars are through stellar mergers or single-star evolution. Indeed, these two channels can account for single Be stars whose space velocities are consistent with those of typical early-type Galactic stars. However, a population of runaway Be stars has also been identified — single Be stars exhibiting significantly higher space velocities (e.g., \citealp{2018MNRAS.477.5261B,2022ApJS..260...35W,2023A&A...679A.109C,2024BSRSL..93..103C}). Such anomalous kinematics suggest an alternative formation pathway. A widely accepted explanation is the binary supernova scenario: in a massive binary system, the primary undergoes a core-collapse supernova (CCSN), releasing the secondary — which may have already been spun up into a Be star via accretion — as a high-velocity runaway star (e.g., \citealp{2015A&A...584A..11L,2018MNRAS.477.5261B,2020MNRAS.497.5344E,2021AN....342..553L,2022ApJ...936..112D}). In this work, we propose another possible origin for single Be stars: the ejection of a companion star following a Type Ia supernova (SN Ia). 

A SN Ia is widely understood to result from the thermonuclear explosion of a carbon-oxygen (CO) WD in a close binary system, with the WD approaching the Chandrasekhar mass limit \citep{1960ApJ...132..565H,1997Sci...276.1378N,1997NuPhA.621..467N,2012A&A...546A..70T}. Recently, an alternative channel has been proposed, in which a He star may also undergo a SN Ia explosion. In this scenario, the He star develops a highly degenerate oxygen-neon (ONe) core, and explosive oxygen burning is triggered by the ignition of residual central carbon \citep{2008NewAR..52..381P,2008ASPC..391..359W,2020A&A...635A..72A,2022A&A...668A.106C,2023MNRAS.526..932G}. If such a He star resides in a binary system, the subsequent SN Ia explosion would completely disrupt the binary, ejecting the companion star. Notably, He stars with masses in the range of $1.5-6.6\ M_\odot$ can undergo He shell burning, driving a mass transfer phase known as Case BB or BC mass transfer \citep{1986A&A...167...61H,2002MNRAS.331.1027D,2015MNRAS.451.2123T,2021ApJ...920L..36J,2024A&A...691A.214Q}. If the companion is a MS star, accretion during this phase could spin it up, transforming it into a classical Be star. In this scenario, should the He star eventually undergo a SN Ia explosion, the end product would be a single Be star.

In this work, we systematically investigate the formation of single Be stars as SN Ia remnants through detailed binary evolutionary simulations and characterize the properties of the resulting single Be stars. The remainder of this paper is as follows. Section \ref{sect:method} describes the simulated methods. The main results are presented in Section \ref{sect:result}, and the discussions and conclusions are provided in Section \ref{sect:disc}.

\section{Method}
\label{sect:method}

In this work, we first adopt the detailed binary evolution code to get the parameter spaces of single Be stars as SN Ia remnants (hereafter, the SN Ia channel). Subsequently, we perform rapid population synthesis to get the Galactic single Be star populations. The detailed methods are described below.

\subsection{Detailed Binary Evolution Simulations}

This study employs the MESA code (version 12115; \citealt{2011ApJS..192....3P,2013ApJS..208....4P,2015ApJS..220...15P,2018ApJS..234...34P,2019ApJS..243...10P}) to simulate binary evolution and examine the formation of a Be star from the SN Ia channel. Specifically, we use the MESA's binary module to simultaneously evolve both components within each system. The main input parameters are described below.

\subsubsection{Stellar-model physics}

A solar metallicity of $Z = 0.02$ is adopted. The mixing length parameter is set to $\alpha_{\text{mlt}} = 1.5$ \citep{1958ZA.....46..108B,1991A&A...252..669L,2008A&A...486..951G}, and semi-convection is modelled with an efficiency factor of $\alpha_{\text{sc}} = 0.01$ \citep{1989A&A...224L..17L,1991A&A...252..669L}. Rotation mixing parameters are adopted following \citet{2000ApJ...544.1016H}. The transport of composition and angular momentum due to rotation includes the effects of Eddington–Sweet circulation, secular and dynamical shear instabilities, as well as the Goldreich-Schubert-Fricke instability, with an efficiency factor $f_{\rm c} = 1/30$ \citep{2016A&A...588A..50M}. The overshooting is implemented as a step function up to 0.335 times the pressure scale height. Stellar wind mass-loss rates are modeled following \citet{2011A&A...530A.115B}. Specifically, for MS stars, the mass-loss rates are taken from \citet{2001A&A...369..574V}, while for He stars, they are calculated using the prescription of \citet{1995A&A...299..151H}, scaled by a factor of 0.1.

\subsubsection{Binary grid}

We begin our simulations by evolving a series of He star + MS star binaries through the SN Ia channel, resulting in the formation of a Be star. In such binary systems, the He star first fills its Roche lobe and undergoes mass transfer. As the MS star accretes mass, it gains both mass and angular momentum, causing it to spin up. 

Our implementation of binary evolution in MESA follows the main inputs detailed by \citet{2022A&A...659A..98S}
, which we summarize below. The mass transfer rate is computed according to the prescription of \citet{2016A&A...588A..50M}. The accretion of angular momentum follows the prescription of \citet{2013ApJ...764..166D}. We adopt the model of \citet{2009MNRAS.396.1699S} for accretion onto rapidly rotating stars, in which the accretion rate is reduced by a factor of ($1-\Omega/\Omega_{\text{crit}}$). Here, $\Omega$ and $\Omega_{\text{crit}}$ denote the angular velocity of the star and its critical value, respectively. When $\Omega = \Omega_{\text{crit}}$, the star will no longer accrete mass, and the remaining material is ejected as an isotropic wind carrying the accretor’s specific orbital angular momentum, $j_{\text{iso}}=(M_1/M_2)(J_{\text{orb}}/M_\text{T})$ (see also \citealt{2005A&A...435.1013P,2009A&A...507L...1D}). Here, $M_1$ and $M_2$ are the donor and accretor masses, respectively, while $M_\mathrm{T} \equiv M_1 + M_2$ and $J_\mathrm{orb}$ represent the orbital angular momentum.

The initial condition assumes synchronized rotation for both components at zero-age main-sequence (ZAMS). Time-dependent tidal forces are then modeled following \citet{2008A&A...484..831D}, adopting the synchronization timescale of \citet{1997A&A...322..320Z} dynamical tide model, which is suitable for radiative-envelope MS stars. In this work, a star is typically classified as a Be star when its rotational angular velocity reaches $0.7\ \Omega_{\rm crit}$. 

A comprehensive parameter space is constructed by evolving models with He star masses ranging from $2.5$ to $2.7\ M_\odot$ in steps of $0.1\ M_\odot$, and MS star (Be star progenitor) masses ranging from $2.0$ to $20.0\ M_\odot$ in steps of $2.0\ M_\odot$. The initial orbital periods range from ${\rm log_{10}}(P\rm_{orb}) = –0.5$ to $3.1$ days (this covers all possible initial orbital periods for forming Be stars), with an interval of $0.1$. We assume that a mass-transfer rate exceeding $0.1\,M_\odot\rm yr^{-1}$ triggers dynamically unstable Roche lobe overflow (e.g., \citealt{2023ApJS..264...45F}). This critical threshold is never reached in our binary evolutionary examples, because mass transfer from a He star is typically more stable (e.g., \citealt{2024ApJS..274...11Z,2026ApJ...999..253Z}).


For He stars with masses between $2.5\ M_\odot$ and $2.7\ M_\odot$, degenerate carbon shell burning may occur. Simulating this phase is computationally expensive; therefore, our code stops when degenerate carbon shell burning occurs. The final evolutionary outcomes of He stars are adopted from \citet{2023MNRAS.526..932G}, who evolved He star binaries until explosive oxygen burning was triggered, and provided the initial parameter ranges (initial He star mass and initial orbital period) that yield SNe Ia (see also \citealt{2024MNRAS.530.4461G}). It should be noted that only the case of a NS accretor with a mass of $1.35\ M_\odot$ was considered in their work, whereas in our work, the accretor masses range from $2-20\ M_\odot$. We have examined our binary simulation results and find that the final He star mass depends only weakly on the accretor mass, with a variation of less than $0.03\ M_\odot$ when the He star fills its Roche lobe at a similar evolutionary stage (see also Figure \ref{Fig2} below). For simplicity, we ignore this influence of the accretor mass and determine the final product of He stars via linear interpolation based on the SN Ia boundaries from \citet{2023MNRAS.526..932G}. Based on the above methods, we could get the parameter spaces of single Be stars from the SN Ia channel. 


\subsection{Binary Population Synthesis Simulation}
\label{sect:bpssimulation}

Based on the calculated parameter grid, we then perform rapid population synthesis to obtain the Galactic population of single Be stars from the SN Ia channel. The specific methods and inputs are described below. First, we adopt the rapid population synthesis code BSE \citep{2000MNRAS.315..543H,2002MNRAS.329..897H,2006MNRAS.369.1152K} to evolve a set of primordial binaries, each consisting of two ZAMS stars. In each binary system, both stars evolve simultaneously. The more massive star fills its Roche lobe first, initiating a mass transfer phase that may be either stable or unstable. As a result of this mass transfer event, the primary star loses its hydrogen-rich envelope and evolves into a He star. Throughout this process, its less massive companion remains on the MS stage. We mainly determine whether a single Be star can form by checking if the simulated He star + MS star binaries fall within the pre-computed parameter grid. Subsequently, we linearly interpolate the binary parameters of He star + MS star binaries within the pre-calculated grids from binary evolution calculations, thereby getting the physical quantities of Be stars formed through the SN Ia channel. The lifetime of a Be star is commonly assumed to be its MS lifetime, consistent with the observational fact that most Be stars exhibit spectra typical of MS stars (e.g., \citealt{1982A&A...109...48M, 2006MNRAS.371..252L}). The progenitors of He star + MS star binaries may experience the CE phase, which is one of the most uncertain physical processes in astrophysics. In this work, we adopt the standard energy mechanism to simplify this process, and in the fiducial model, the CE ejection efficiency is assumed to be $1$. The main inputs of the binary population synthesis (BPS) are introduced as follows.

The initial mass function (IMF) of the primordial primary star is taken from \citet{1979ApJS...41..513M} and \citet{1989ApJ...347..998E}, in which $M_{1,\mathrm{i}}$ ranges from $0.1$ to $100\ M_\odot$. We generate the primary mass with the formula of \citet{1989ApJ...347..998E},
\begin{equation}
M_{1, \mathrm{i}}=\frac{0.19 X}{(1-X)^{0.75}+0.032(1-X)^{0.25}},
\end{equation}
where $X$ is a random number uniformly distributed between $0.4$ and $0.999992$. In this work, we only evolve primary stars with masses between $1$ and $50\ M_\odot$. A constant mass-ratio distribution is taken, i.e., $n(1/q_\mathrm{i}) = 1$ (for $q_\mathrm{i} > 1$), in which $q_\mathrm{i} \equiv M_{1,\mathrm{i}}/M_{2,\mathrm{i}}$, where $M_{1,\mathrm{i}}$ and $M_{2,\mathrm{i}}$ are the initial masses of the primordial primary and secondary, respectively \citep{1992ApJ...401..265M}. The initial separation distribution takes from \citet{1998MNRAS.296.1019H}:
\begin{equation}
an(a)=\left\{\begin{array}{lrl}
{\alpha }_{\rm sep}{\left ({a}/{{a}_{0}}\right )}^{m}, & a & \leq a_0, \\
{\alpha }_{\rm sep}, & a_0 & \leq a \leq a_1,
\end{array}\right.
\end{equation}
with $\alpha_{\rm sep} \approx 0.07$, $a_0 = 10\ R_\odot$, $a_1 = 5.75\times10^6$, $R_\odot = 0.13\ \mathrm{pc}$, and $m = 1.2$. We simply take a constant star formation rate over the last 15 Gyr at $S = 3\ M_\odot /{\rm yr}$ \citep{1978A&A....66...65S,2006Natur.439...45D,2010ApJ...710L..11R}. 

\section{Results}
\label{sect:result}
\subsection{Parameter Space for Single Be Stars}

\begin{figure}
\centering
\includegraphics[width=\textwidth, angle=0]{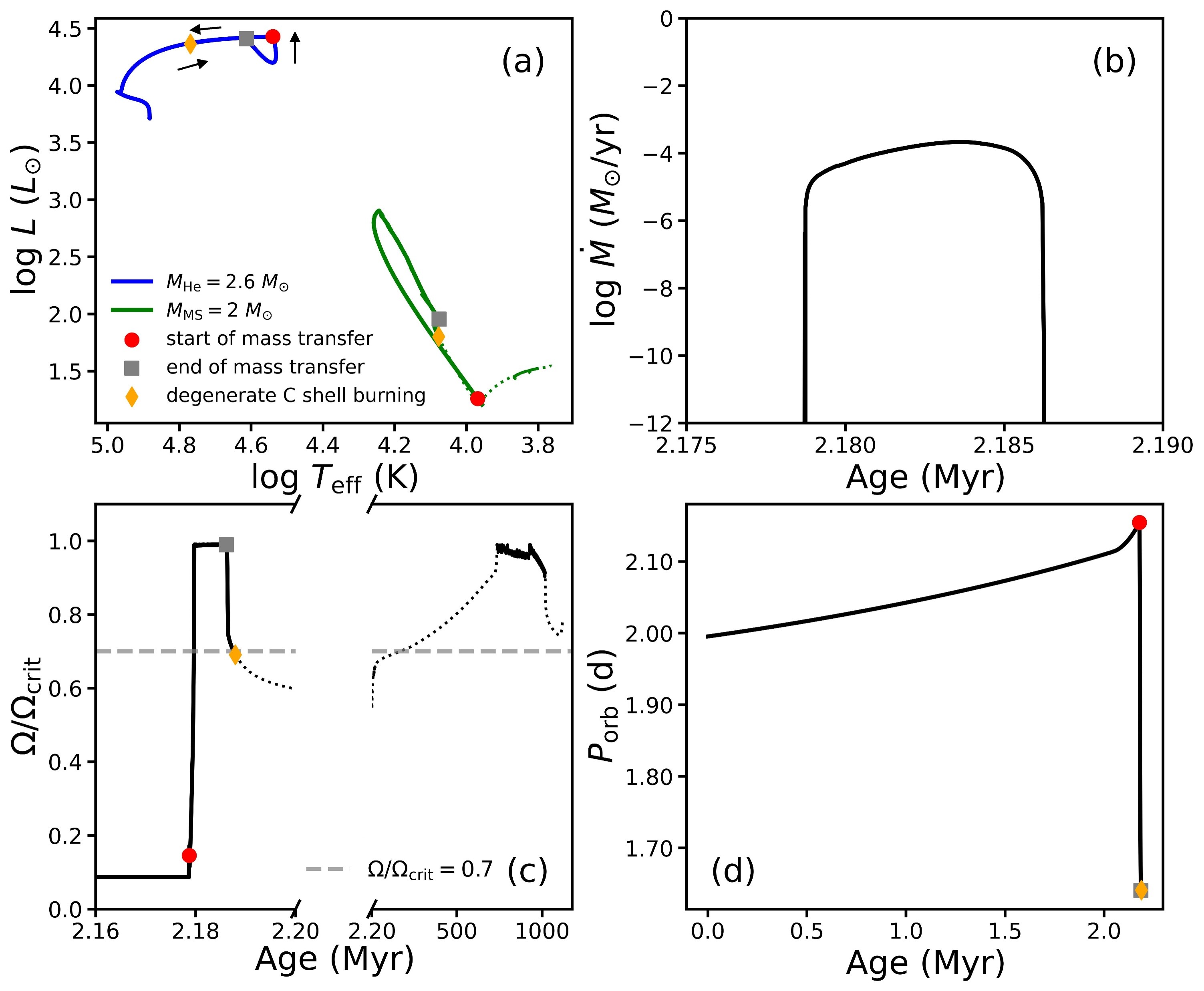}
\caption{Panel (a) display the Hertzsprung–Russell diagrams for the components of a binary system: a $2.6\ M_\odot$ He star and a $2\ M_\odot$ MS star in a $2$-day orbit. Panel (b) shows the evolution of the mass transfer rate with age. Panel (c) shows the time evolution of $\Omega/\Omega_{\rm crit}$ for the MS star, with a grey dashed line marking the value of $0.7$. The dotted lines in panels (a) and (c) represent the subsequent evolution of the single Be star after the SN Ia explosion. Panel (d) presents the evolution of the orbital period of the MS star as a function of time. In all panels, a red circle marks the onset of mass transfer, a grey square marks its end, and an orange diamond marks the onset of degenerate carbon shell burning (the simulation endpoint).}
\label{Fig1}
\end{figure}

Figure \ref{Fig1} illustrates a typical evolutionary pathway for the formation of a single Be star. The initial binary system consists of a $2.6\ M_\odot$ He star and a $2\ M_\odot$ MS star in a $2$-day orbit. Panel (a) shows the Hertzsprung–Russell diagrams for the He star and the MS star. Panel (b) displays the evolution of the mass transfer rate with age. Panel (c) displays the time evolution of the rotational velocity ratio $\Omega/\Omega_{\mathrm{crit}}$ for the MS star, while panel (d) presents the evolution of the orbital period. The onset and end of mass transfer are indicated by red circles and grey squares, respectively. We terminate the binary simulations at the onset of degenerate carbon shell burning in the He star. To assess whether the resulting single Be star can maintain a high rotation velocity, we continue its evolution as a single star using the stellar parameters at the end of the binary simulation. The single-star calculation is stopped at the end of the MS phase, defined here as the onset of He core burning. The corresponding evolutionary tracks are shown as dotted lines in panels (a) and (c).

In panel (a), the He star evolves first, fills its Roche lobe, and initiates mass transfer during its He shell burning phase (i.e., Case BB mass transfer). At the onset of mass transfer, the He star has a mass of $2.42\ M_\odot$ with a CO core mass of $1.2\ M_\odot$. After the mass-transfer phase, the He star mass decreases to approximately $1.54\ M_\odot$ with a CO core mass of $1.24\,M_\odot$ The mass transfer rate as a function of evolutionary age is shown in Panel (b); the mass transfer phase lasts about $7.5 \times 10^3\ \rm yr$. The onset of degenerate carbon shell burning is marked by the orange diamond. Following the simulations of \citet{2023MNRAS.526..932G,2024MNRAS.530.4461G}, such a He star subsequently undergoes explosive oxygen burning, leading to a SN Ia explosion. During mass transfer, the accretor is driven out of thermal equilibrium and initially expands (e.g., \citealt{2024MNRAS.531L..45Z}). As a result of angular momentum accretion, the rotational velocity of the MS star increases sharply (panel c), eventually reaching critical rotation. Once this critical rotation is achieved, the star ceases to accrete further material and gradually contracts to re-establish thermal equilibrium, triggering a corresponding decrease in its radius and luminosity (panel a).

Panel (c) shows that the MS star rapidly becomes a Be star during the accretion phase. Following the end of mass transfer, the value of $\Omega/\Omega_{\rm crit}$ declines. This decrease is because the star, still out of thermal equilibrium, begins to contract towards its equilibrium state. This contraction causes an increase in the critical rotation velocity $\Omega_{\rm crit}$, thereby reducing the $\Omega/\Omega_{\rm crit}$. It should be noted that the inner regions of the MS star retain significant spin angular momentum. During subsequent evolution, this angular momentum can be redistributed from the interior to the envelope, potentially causing the star to re-enter a Be star phase (e.g., \citealt{2019MNRAS.485.3661F}). We find that for the majority of its MS lifetime, the value of $\Omega/\Omega_{\rm crit}$ remains above $0.7$. Therefore, the accretor can be regarded as a single Be star following the SN Ia explosion.

The panel (d) illustrates the evolution of the orbital period of the companion star. Before the onset of mass transfer, the orbital period increases mainly due to stellar wind mass loss, which results in an increase in the orbital separation. At the onset of mass transfer, the orbital period decreases due to the angular momentum loss caused by unprocessed material, leading to orbital shrinkage. The time between the end of mass transfer and the onset of degenerate carbon shell burning is very short, approximately $1.68 \times 10^3\ \rm yr$, during which the orbital period remains nearly unchanged.

\begin{figure}
\centering
\includegraphics[width=\textwidth, angle=0]{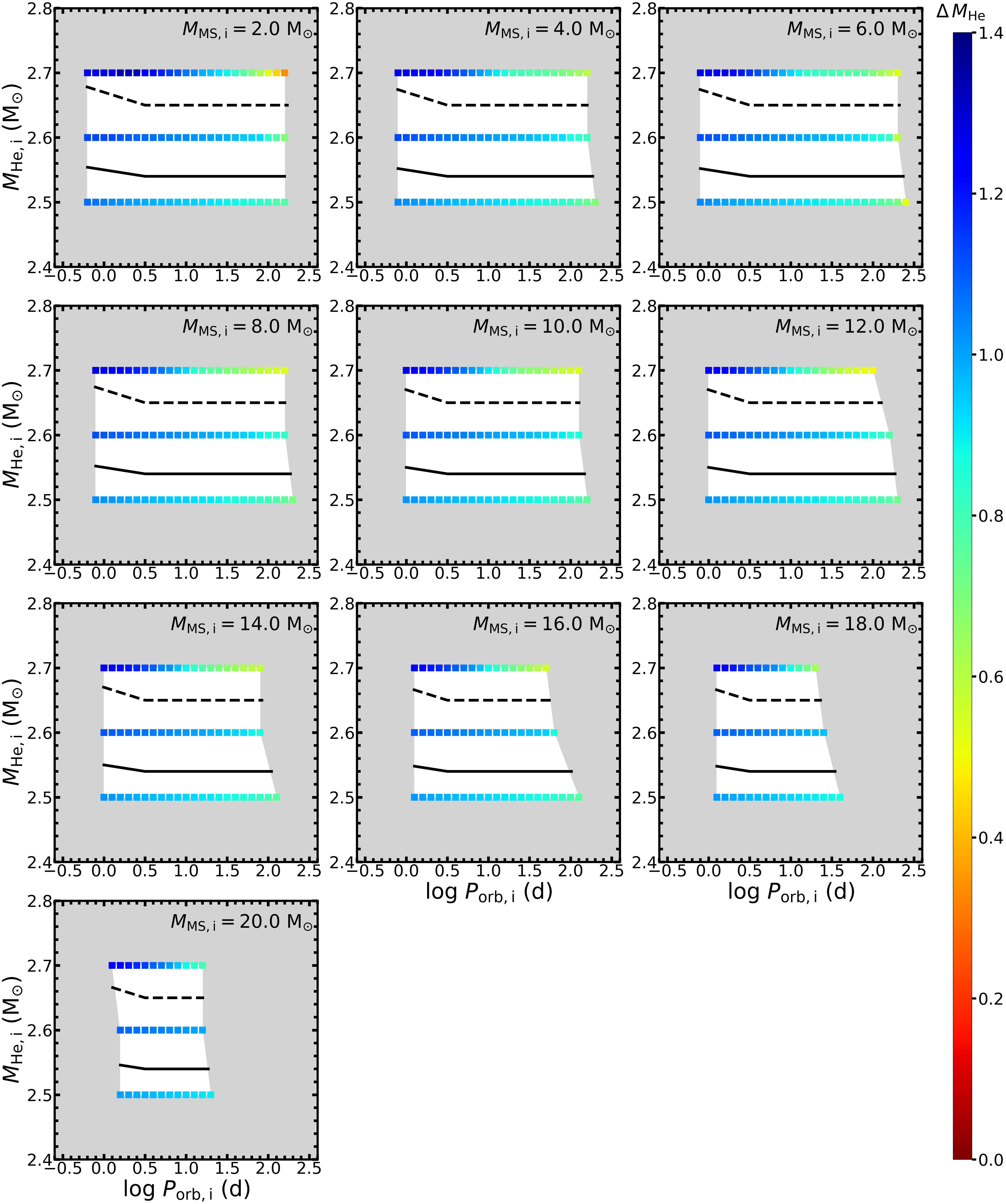}
\caption{$P_{\mathrm{orb,i}}-M_{\mathrm{He,i}}$ parameter space for different initial MS masses. The colorbar is the mass loss by the He star during mass transfer, defined as $\Delta M_\mathrm{He} = M_{\mathrm{He,i}} - M_{\mathrm{He,f}}$. Systems above the black dashed line will undergo an ECSN or FeCCSN, while those below the black solid line will produce a WD. The region between the dashed and solid lines corresponds to primordial binaries that produce a SN Ia.}
\label{Fig2}
\end{figure}

Building upon this evolutionary scenario, we map out the parameter space for Be star formation via the SN Ia channel, as illustrated in Figure \ref{Fig2}. The colored squares correspond to initial binary parameters for which Be stars are formed. According to the upper and lower boundaries for SN Ia from \citet{2023MNRAS.526..932G}, the initial mass of He stars is constrained to the range of $2.5-2.7\,M_\odot$ in this work. The colorbar corresponds to the mass change of the He star during the mass transfer process, defined as $\Delta M_{\rm He} = M_{\rm He,i}-M_{\rm He,f}$. Here, $M_{\rm He,i}$ is the initial mass of the He star, and $M_{\rm He,f}$ is its mass after the mass transfer. \citet{2023MNRAS.526..932G,2024MNRAS.530.4461G} have established a well-defined boundary between SN Ia and electron-capture supernovae (ECSN)/iron core-collapse 
supernovae (Fe CCSN) outcomes, indicated by the black dashed line. Initial binary parameters above this line correspond to He stars that will undergo an ECSN or Fe CCSN. The black solid line represents the boundary between WD formation and SN Ia: for parameters below this line, the final He star mass evolves into a WD. Single Be stars form once the initial binary parameters fall into the regions between the dashed and solid lines.


\subsection{Population Properties of Single Be stars}

\begin{figure}
\centering
\includegraphics[width=0.8\textwidth, angle=0]{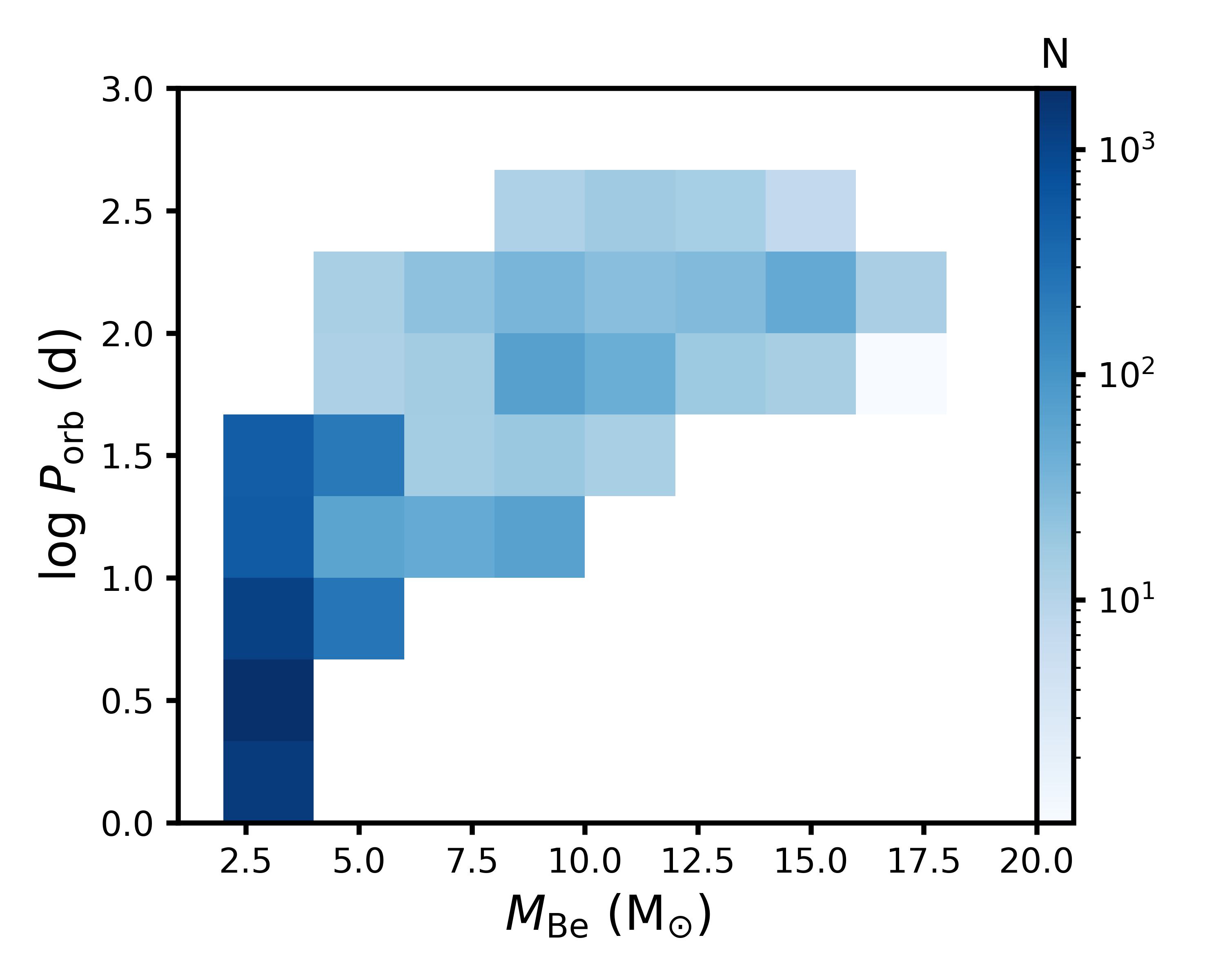}
\caption{Simulated distributions of Galactic single Be stars in the $M_{\mathrm{Be}}-P_{\mathrm{orb}}$ plane at the moment of SN Ia explosion. Each panel displays a $10 \times 10$ parameter matrix, with pixel color representing the number of single Be stars in each bin.}
\label{Fig3}
\end{figure}

\begin{figure}
\centering
\includegraphics[width=0.8\textwidth, angle=0]{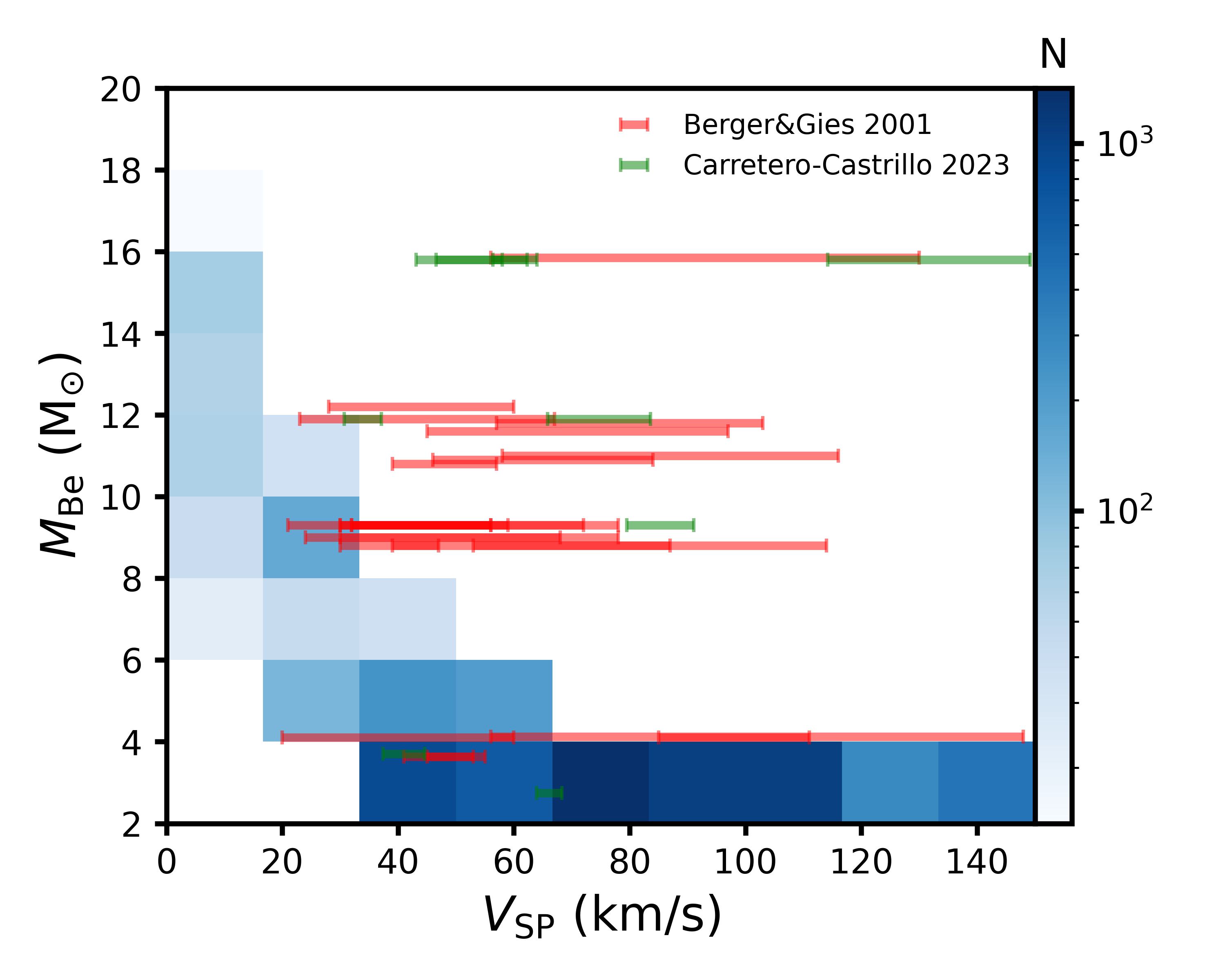}
\caption{Similar to Figure \ref{Fig3}, but for the simulated distributions of Galactic single Be stars in the $V_{\mathrm{SP}}$–$M_{\mathrm{Be}}$ plane. The red and green lines are the observational Be stars from \citet{2001ApJ...555..364B} and \citet{2023A&A...679A.109C}, respectively. The low-mass Be stars ($< 5\ M_\odot$) may originate from the SN Ia channel explored in this work, while the high-mass Be stars ($\ge 5\ M_\odot$) probably come from the other channels, such as the ejected companion stars of CCSN.}
\label{Fig4}
\end{figure}

\begin{figure}
\centering
\includegraphics[width=0.8\textwidth, angle=0]{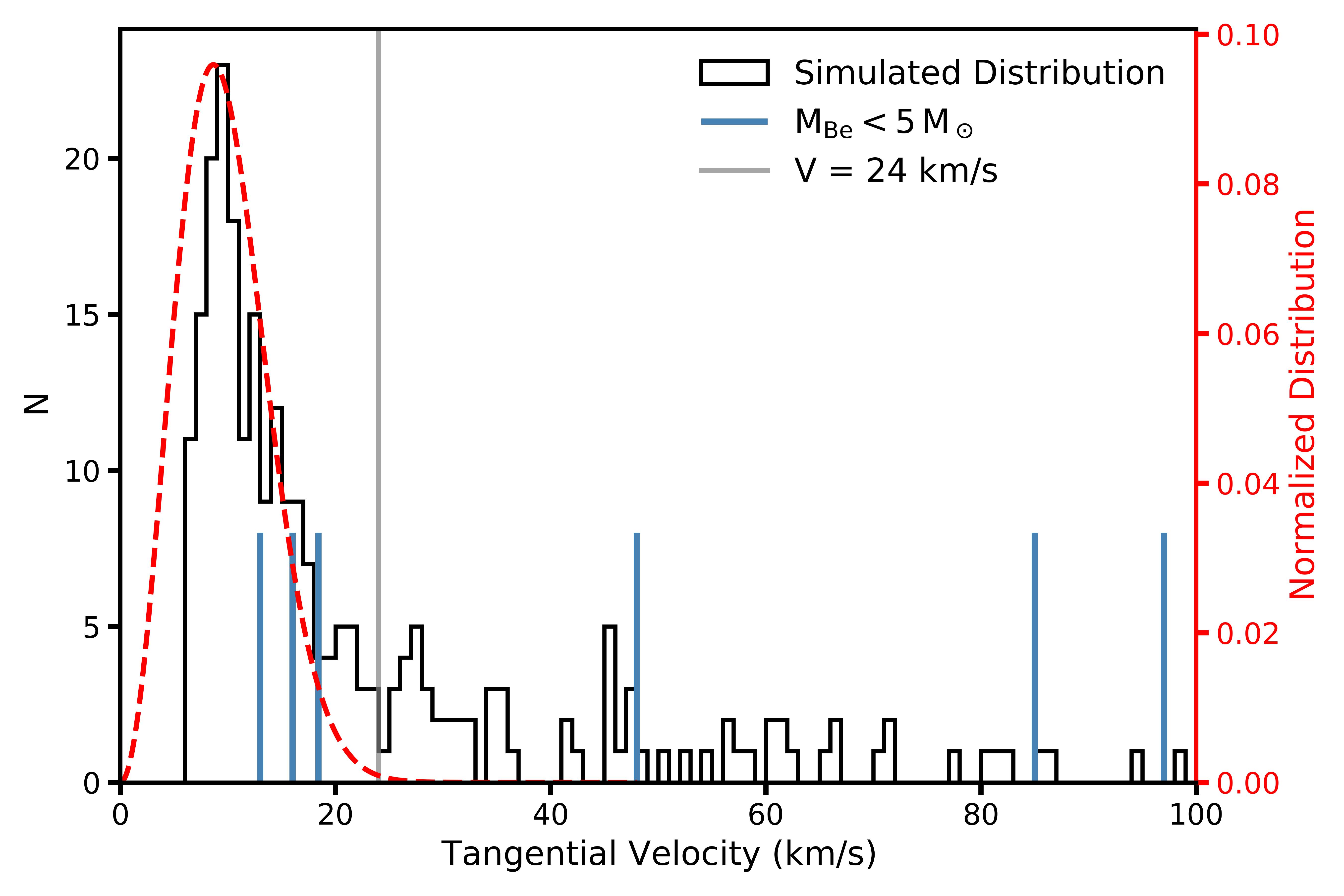}
\caption{Tangential velocity distributions for simulated Galactic single Be stars. The black histogram shows the results from our simulations. The red curve shows the Maxwellian fit to the observed classical Be stars from \citet{2022ApJS..260...35W}. The grey line indicates the $24\ \rm km/s$ tangential velocity threshold for defining a runaway star. The blue lines represent the median tangential velocity of the observed low-mass Be stars ($< 5\ M_\odot$) compiled from the \citet{2001ApJ...555..364B} and \citet{2023A&A...679A.109C}.}
\label{Fig5}
\end{figure}

Based on the parameter spaces depicted in Figure \ref{Fig2}, we performed a rapid population synthesis to estimate the population of single Be stars in the Galaxy. We evolved $10^7$ primordial binaries with initial parameters following the assumptions described in Section \ref{sect:bpssimulation}. This yields $86810$ He star + MS star binaries, of which $343$ (approximately $0.4\%$) fall within the single-Be-star parameter space identified in Figure \ref{Fig2}. After convolving with the star formation history, we derive the resulting population of single Be stars in the Galaxy. Our results indicate that the total number of single Be stars produced via the SN Ia channel is approximately $6.6 \times 10^3$. For comparison, \citet{2014ApJ...796...37S} simulated the formation of single Be stars as ejected companions of CCSN, predicting numbers of around $4000$–$20000$. This shows that the contribution from the SN Ia channel is comparable to that from the CCSN channel.

Figure \ref{Fig3} presents the Galactic density distributions of the binary parameters, including the Be star and orbital period at the moment of SN Ia explosion. The Be star masses span $2-20\ M_\odot$ with a characteristic peak at $\sim$ $3\ M_\odot$. Beyond this peak, the distribution falls off monotonically toward higher masses, as lower-mass stars have longer MS lifetimes (similar outcome observed in relevant population synthesis works; e.g., \citealp{2014ApJ...796...37S,2025ApJ...987..210B,2026ApJ...996L..42L}).

We find that for massive Be stars ($\gtrsim 6\ M_\odot$), the corresponding orbital periods after mass transfer are typically longer than $10\ \rm d$, whereas for lower-mass Be stars, many systems exhibit much shorter orbital periods. This trend can be understood as follows. In the case of massive Be star progenitors, the progenitor binaries generally have small mass ratios (primary/secondary), which typically favors stable mass transfer. As a result, the resulting He star + MS star systems tend to have longer orbital periods, leading also to long orbital periods in the subsequent Be star binaries. Conversely, for progenitor binaries with a low-mass secondary, the first mass transfer event is more likely to be unstable, potentially driving the system into a CE phase. This process typically produces He star + MS star systems with initially short orbital periods, which then remain relatively short after the subsequent mass transfer between the He star and the MS star.

Assuming that the SN Ia occurs instantaneously and the SN ejecta has no effect on the companion star, the Be star then has a space velocity equal to the orbital velocity at the moment of SN explosion. Figure \ref{Fig4} displays the distributions of the masses and the space velocities for Galactic Be stars at the moment of SN Ia explosion. According to Kepler’s laws, the orbital velocity can be calculated from the pre-explosion binary parameters: the masses of both stars and the orbital period. This velocity corresponds to the resulting space velocity of the Be star. Here, the relevant parameters of Be stars are taken from \citet{2001ApJ...555..364B} and \citet{2023A&A...679A.109C}. From these samples, we select only Be stars that do not show clear binary signals in the observations and treat them as single Be stars. Using the observed spectral types, we determined the stellar masses and present them in Figure \ref{Fig4} \citep{1981A&AS...46..193H}. This distribution, similar to that shown in Figure \ref{Fig3}, indicates that more massive Be stars have lower space velocities, while less massive ones have higher velocities. A comparison with observations indicates that the lower-mass end of the observed Be star sample can be reproduced by our theoretical simulations, suggesting that these stars may originate from the SN Ia channel. In contrast, higher-mass Be stars in the sample are likely formed through different channels, such as ejected companion stars of CCSN \citep{2014ApJ...796...37S,2026ApJS..283...59A}.

\citet{2014ApJ...796...37S} performed detailed simulations of single Be stars formed via CCSN, identifying two distinct evolutionary pathways: the MS star channel (hereafter MS star CCSN channel) and the He star channel (hereafter He star CCSN channel). In the MS star CCSN channel, the predicted number of single Be stars ranges from about 2000 to 20000 depending on accretion efficiency, and the resulting Be stars typically have masses exceeding $8\ M_\odot$ with pre-CCSN orbital periods longer than $10$ days, corresponding to space velocities below $100\,\rm km/s$ (see also \citealt{2021ApJ...908...67S}). The He star CCSN channel shares similarities with the channel studied in our work, as the physical properties of the resulting single Be stars (e.g., mass and space velocity) are similar between these two channels. However, a key distinction lies in their progenitors: He stars with masses greater than approximately $2.9\ M_\odot$ undergo CCSN (see \citealt{2024MNRAS.530.4461G}), while those in the range of about $2.5$ to $2.7\ M_\odot$ produce SNe Ia. The mass distribution of He stars, sensitive to the IMF, leads to a predominance of lower-mass He stars. \citet{2014ApJ...796...37S} estimated that the number of single Be stars from the He star CCSN channel is 2500 to 4000, whereas our work finds that the SN Ia channel yields about 6600 such stars. Comparing the characteristics from different channels, we draw two main conclusions. First, compared to the MS star CCSN channel, our SN Ia channel produces a larger number of low-mass single Be stars. Second, while the orbital properties (e.g., space velocities and orbital periods) of Be stars from our SN Ia channel and the He star CCSN channel are likely similar due to comparable binary progenitor parameters, our channel yields a significantly higher number of single Be stars, primarily due to the more abundant low-mass He star progenitors favored by the IMF.

In a recent study, \citet{2022ApJS..260...35W} derived the peculiar tangential velocities of 323 classical Be stars using data from the LAMOST Medium Resolution Survey, providing an important observational reference for investigating their physical properties. To compare with their results, we derived the tangential velocity distribution for single Be stars from the SN Ia channel by randomly sampling the angle between the space velocity vector and the line of sight. As shown in Figure \ref{Fig5}, \citet{2022ApJS..260...35W} found that the observed tangential velocities for most Be stars follow a Maxwellian distribution (red dashed line). They further defined runaway stars as those with tangential velocities exceeding the value at $1\%$ of the peak of the fitted Maxwellian distribution, corresponding to a threshold of $24\ \rm km/s$ (grey line). The tangential velocity distribution of observed low-mass Be stars ($< 5\ M_\odot$) is also plotted for comparison. Our simulated distribution for Be stars from the SN Ia channel peaks around $10\ \rm km/s$, consistent with the observational peak. However, it should be noted that the sample of \citet{2022ApJS..260...35W} likely includes Be stars from all formation channels (such as merger channel and single-star channel), meaning that Be stars with tangential velocities near $10\ \rm km/s$ cannot be attributed to a specific channel based on kinematics alone. We predict that approximately $22\%$ of Be stars formed through the SN Ia channel may exceed the runaway threshold and thus be classified as runaway stars.

\section{Discussions and Conclusions}
\label{sect:disc}

In this study, we propose a new formation channel for single Be stars, in which the Be star is the ejected companion of a SN Ia. Previous studies have indicated that surviving companions of SN Ia may include hot subdwarfs, WDs, and NSs, among others (e.g., \citealp{2013ApJ...773...49P,2018MNRAS.479..192K,2019ApJ...886...99L,2019MNRAS.482.5651M,2000ApJS..128..615M,2001ApJ...550L..53C,2019NewAR..8501523R,2021MNRAS.507.4603M,2022ApJ...933...38R,2023ApJ...945....7G,2023RAA....23h2001L}). Here, we suggest that a Be star could represent another possible type of surviving companion. However, observationally confirming a Be star as the surviving companion of a SN Ia remains challenging. After the SN, the companion’s atmosphere may become polluted by SN ejecta. As a result, if Be stars indeed originate from the SN Ia channel, their spectra could exhibit an enhancement in iron-peak elements. Nevertheless, such enhancement may be difficult to distinguish from abundance anomalies caused by radiative levitation.

The effect of metallicity represents an important factor in the formation and evolution of Be stars, especially given that Be stars are observed in environments with different metallicities, such as the Small and Large Magellanic Clouds \citep{1999A&AS..134..489K,2009IAUS..256..349M,2013MNRAS.435.3103I,2022A&A...667A.100S,2025A&A...698A..38B}. In the current work, we adopt solar metallicity for our models. The expected influence of metallicity on our results is as follows. As the metallicity decreases, both the initial orbital periods of the binaries and the radii of the He stars become smaller \citep{2022A&A...668A.106C,2023A&A...671A.134A}. The smaller stellar radii lead to shorter orbital periods at the onset of mass transfer, which in turn results in shorter orbital periods for the resulting Be stars \citep{2010A&A...515A..88W}. This leads to higher orbital velocities and thus higher tangential velocities of Be stars. In contrast, higher metallicity would produce the opposite effect, potentially narrowing the parameter space and shifting the tangential velocity distribution toward lower values. Moreover, the mass range of He stars that give lead to SNe Ia should be metallicity-dependent (see \citealt{2024MNRAS.530.4461G}), which could influence the number and properties of the resulting Be stars.

Our study predicts that approximately $6.6 \times 10^{3}$ single Be stars originate from the SN Ia channel, a population comparable in size to that formed through the CCSN channel. Their mass distribution peaks at $3$–$4\ M_\odot$, and a clear trend emerges: lower-mass Be stars tend to exhibit higher space velocities. We further estimate that about $22\%$ of Be stars from the SN Ia channel possess peculiar tangential velocities exceeding $24\ \rm km/s$, qualifying them as runaway stars. Taken together, these findings highlight the SN Ia channel as a non-negligible contributor to the formation of single Be stars, especially within the runaway population.

\begin{acknowledgements}
We are deeply grateful to the anonymous referee for the insightful comments, which have significantly improved the quality of this work. We thank Yunlang Guo for helpful assistance in data analysis and insightful suggestions. This work is supported by the Strategic Priority Research Program of the Chinese Academy of Sciences (grant Nos. XDB1160303, XDB1160201, XDB1160000), the National Natural Science Foundation of China (NSFC, grant Nos. 12125303, 12525304, 12288102, 12473034, 12273105, 11703081, 11422324, 12073070, 12333008, 12422305, and 12473033), the CAS Project for Young Scientists in Basic Research (YSBR-148), the National Key R$\&$D Program of China (grant Nos. 2021YFA1600403, 2021YFA1600400), the Yunnan Revitalization Talent Support Program-Science $\&$ Technology Champion Project (No. 202305AB350003), the International Centre of Supernovae (ICESUN), Yunnan Key Laboratory of Supernova Research (Nos. 202302AN360001, 202201BC070003), Yunnan Fundamental Research Projects (No. 202401AT070139), the New Cornerstone Science Foundation through the XPLORER PRIZE, the Young Talent Project of Yunnan Revitalization Talent Support Program, and the Key Research Program of Frontier Sciences of CAS (No. ZDBS-LY-7005). The authors gratefully acknowledge the "PHOENIX Supercomputing Platform" jointly operated by the Binary Population Synthesis Group and the Stellar Astrophysics Group at Yunnan Observatories, Chinese Academy of Sciences.
\end{acknowledgements}

\bibliographystyle{raa}
\bibliography{bibtex}

\label{lastpage}

\end{document}